\RequirePackage{lineno}
\documentclass[prd,twocolumn,nofootinbib,superscriptaddress,floatfix]{revtex4}
\usepackage{cancel}
\usepackage{graphicx}
\usepackage{epsfig}
\usepackage{amsmath,amsfonts,amssymb}
\usepackage{t1enc}

\newcommand{\vl}{V_L}
\newcommand{\vR}{V_R}
\newcommand{\gl}{g_L}
\newcommand{\gr}{g_R}
\newcommand{\R}{\text{Re}}
\newcommand{\I}{\text{Im}}

\begin{document}

\title{New directions for top quark polarization in the $t$-channel process}

\author{J.A. Aguilar-Saavedra}
\affiliation{Departamento de F\'{\i}sica Te\'orica y del Cosmos, Universidad de Granada,
 E-18071 Granada, Spain.}
\affiliation{PH-TH Department, CERN, CH-1211 Geneva 23, Switzerland.}
 \author{S. Amor dos Santos}
 \affiliation{LIP, Departamento de F\'\i sica da Universidade de Coimbra, P-3004-516 Coimbra, Portugal.}

\begin{abstract}
We define three orthogonal axes to investigate the top quark polarization in the $t$-channel single top process. We provide expressions for the polarization in these axes in terms of anomalous $Wtb$ couplings. It is found that the polarizations in the two axes orthogonal to the spectator quark axis are very sensitive to an anomalous coupling involving a $\bar b_L \sigma^{\mu \nu} t_R$ dipole term. In particular, an asymmetry based on the polarization normal to the production plane is more sensitive to the imaginary part of this coupling than previously studied observables.
\end{abstract}

\maketitle

\section{Introduction}

Single top quark production at hadron colliders is at present the only source of polarized top quarks. For the $t$-channel process it is well known that, both at the Fermilab Tevatron~\cite{Mahlon:1996pn} and the CERN Large Hadron Collider (LHC)~\cite{Mahlon:1999gz}, top quarks are produced with a high polarization in the direction of the spectator quark in the top quark rest frame. (Here we define the polarization along some axis as twice the expectation value of the spin operator in that axis; the spectator quark in $qg \to q' t \bar b$ is the light quark $q'$.) At the Tevatron, the small statistics and large backgrounds did not allow to test this prediction, but at the LHC the situation has significantly improved. Recently the CMS Collaboration has presented~\cite{CMS:2013rfa} the first measurement of the top polarization in the spectator quark direction, $P = 0.82 \pm 0.12\,\text{(stat)} \pm 0.32
\,\text{(sys)}$, measured at a centre-of-mass energy of 8 TeV. As the top quark is unstable and decays $t \to Wb$, its polarization must be extracted from an analysis of the angular distributions of its decay products in the top quark rest frame -- or by the analysis of related observables. In~\cite{CMS:2013rfa} the angular distribution of the charged lepton $\ell=e,\mu$ from the decay $t \to W b \to \ell \nu b$ was used, because this angular distribution is the most sensitive to the top quark polarization.

In this paper we study the top quark polarization in the $t$-channel process not only in the spectator quark direction but also in two orthogonal axes, hence providing a complete description of the top (anti-)quark polarization. We give for the first time expressions for the polarizations in these axes for a general $Wtb$ effective vertex with anomalous couplings. The sensitivity to these non-standard contributions is also estimated. Previous work~\cite{Espriu:2001vj} investigated the polarization in the spectator quark and helicity axes, and their dependence on one of the three possible anomalous $Wtb$ couplings.

\section{Top quark polarization beyond the spectator quark axis}

Let us review the basics of top polarization in some detail. Introducing a reference system with axes $(x,y,z)$ in the top quark rest frame, the state of an ensemble of polarized top quarks can be described by a density matrix
\begin{equation}
\rho=\frac{1}{2}\left( \! \begin{array}{cc} 1+P_z & P_x-i P_y \\ P_x+i P_y & 1-P_z
\end{array} \! \right) \,,
\end{equation}
where $P_i = 2 \langle S_i \rangle$, with $i=x,y,z$. $\vec P$ is a vector in three-dimensional space, satisfying $|\vec P| \leq 1$ in general, and $|\vec P| = 1$ if, and only if, the top quarks are produced in a pure spin state. The three orthogonal axes can be chosen as follows. The $\hat z$ direction is taken as the direction of the spectator quark three-momentum $\vec p_j$. Then, $\hat y$ is taken orthogonal to $\vec p_j$ and the initial quark three-momentum $\vec p_q$, and $\hat x$ is determined requiring that the coordinate system is right-handed. That is,
\begin{equation}
\hat z = \frac{\vec p_j}{|\vec p_j|} \,,\quad
\hat y = \frac{\vec p_j \times \vec p_q}{|\vec p_j \times \vec p_q|} \,,\quad
\hat x = \hat y \times \hat z \,,
\label{ec:axes}
\end{equation}
with $\vec p_j$ and $\vec p_q$ in the top quark rest frame.
The $\hat z$ direction will be called `longitudinal' hereafter, the $\hat x$ direction `transverse' and the $\hat y$ direction `normal'. This denomination is motivated because in the $2 \to 2$ approximation to $t$-channel production, $q b \to q' t$, the vectors $\vec p_j$ and $\vec p_q$ determine the production plane, hence $\hat x$ is contained in that plane and $\hat y$ is orthogonal to it. Notice that the momentum direction of the initial quark in the $t$-channel process cannot be determined unambiguously in hadronic collisions; we will deal with this apparent difficulty later.
 
In the SM we have polarizations
\begin{align}
& \vec P \simeq (0,0,0.90) && (t) \,, \notag \\
& \vec P \simeq (-0.14,0,-0.86) && (\bar t) \,,
\end{align}
computed for the $2 \to 3$ process $qg \to q' t \bar b$ with the generator {\sc Protos}~\cite{AguilarSaavedra:2008gt}, using CTEQ6L1~\cite{cteq6l1} parton distribution functions.
The longitudinal polarization so calculated agrees very well with the full next-to-leading order (NLO) calculation~\cite{Schwienhorst:2010je}, $P_z = 0.91$, $P_z = -0.86$ for quarks and anti-quarks, respectively. In addition, several kinematical distributions are similar~\cite{Campbell:2009ss} which brings confidence that our results hold with a full NLO computation. Note that $P_x$ is not zero for top antiquarks in the SM because the spectator quark is not the $d$ quark in the leading production process, $dg \to u \bar t b$~\cite{Mahlon:1996pn,Mahlon:1999gz}.

The presence of new physics in $t$-channel single top production dramatically changes these SM predictions. Here, we focus on possible $Wtb$ anomalous couplings. The most general effective $Wtb$ interaction arising from a minimal set of dimension-six effective operators can be written as~\cite{AguilarSaavedra:2008zc}
\begin{eqnarray}
\mathcal{L}_{Wtb} & = & - \frac{g}{\sqrt 2} \bar b \, \gamma^{\mu} \left( \vl
P_L + \vR P_R
\right) t\; W_\mu^- \notag \\
& & - \frac{g}{\sqrt 2} \bar b \, \frac{i \sigma^{\mu \nu} q_\nu}{M_W}
\left( \gl P_L + \gr P_R \right) t\; W_\mu^- + \mathrm{h.c.} \,, \notag \\
\label{ec:lagr}
\end{eqnarray}
where $\vR$, $\gl$ and $\gr$ are the so-called anomalous couplings, which are zero at the tree level in the SM. At present, the most stringent direct limits on them result from the measurement of two asymmetries~\cite{AguilarSaavedra:2006fy} in $t \bar t$ production by the ATLAS Collaboration, $\R\, \vR \in[-0.20,0.23]$, $\R\, \gl \in[-0.14,0.11]$, $\R\, \gr \in[-0.08,0.04]$ with a 95\% confidence level, assuming only one non-zero anomalous coupling~\cite{Aad:2012ky}. (The combination with single top cross sections would improve these limits~\cite{AguilarSaavedra:2011ct}.) Current measurements of the $W$ helicity fractions~\cite{Kane:1991bg} by the ATLAS and CMS Collaborations give less stringent constraints, for example $\R\, \vR \in[-0.31,0.33]$, $\R\, \gl \in[-0.17,0.14]$, $\R\, \gr \in[-0.14,0.02]$ using the measurements in~\cite{CMS:2013pfa} with the full 20 fb$^{-1}$ data sample at 8 TeV.
The individual limits on $\I\,\vR$ and $\I\,\gl$ are expected to be of similar magnitude as those on $\R\,\vR$ and $\R\,\gl$, respectively, since for this size of couplings the quadratic terms $|\vR|^2$, $|\gl|^2$ dominate over the interference ones~\cite{AguilarSaavedra:2006fy}. Current limits on $\I\,\gr$ from the measurement of the normal $W$ polarization in the top quark decay~\cite{AguilarSaavedra:2010nx} are still loose, $\I\, \gr \in[-0.3,0.4]$~\cite{ATLAS:2013ula}. Indirect limits on $\vR$ and $\gl$ from $B$ physics~\cite{Grzadkowski:2008mf,Drobnak:2011wj} are one order of magnitude more stringent than those from top decays, because their $b$-chirality flipping contributions are enhanced. This is not the case for $\gr$, however, and indirect limits from $B$ physics are looser, $\R\, \gr \in[-0.57,0.15]$~\cite{Grzadkowski:2008mf}.
\begin{figure*}[!t]
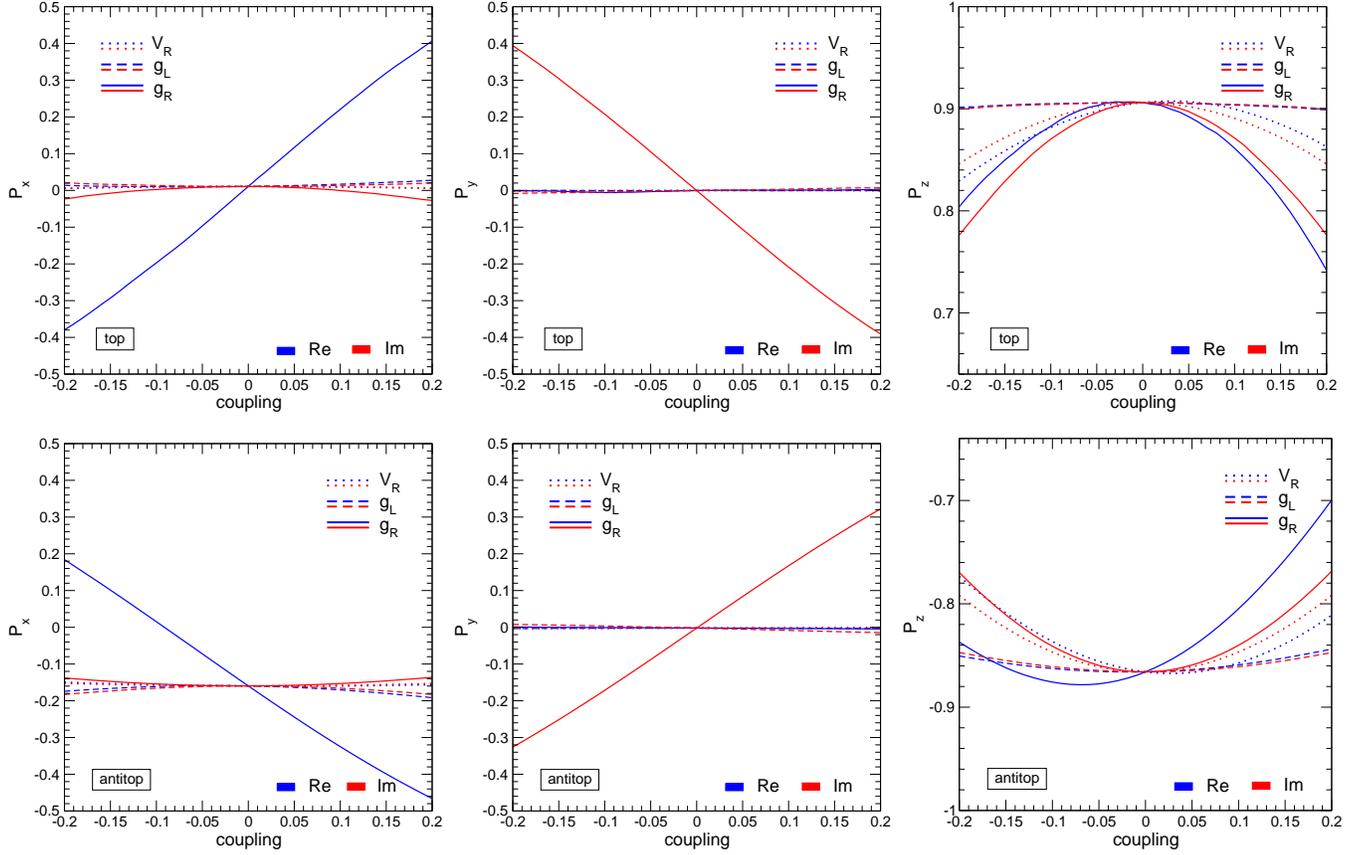

\begin{center}
\begin{tabular}{ccc}
\epsfig{file=fig1a.eps,height=5.5cm,clip=} &
\epsfig{file=fig1b.eps,height=5.5cm,clip=} &
\epsfig{file=fig1c.eps,height=5.5cm,clip=} \\[2mm]
\epsfig{file=fig1d.eps,height=5.5cm,clip=} &
\epsfig{file=fig1e.eps,height=5.5cm,clip=} &
\epsfig{file=fig1f.eps,height=5.5cm,clip=} 
\end{tabular}
\caption{Single top and antitop polarization in the axes defined in Eqs.~(\ref{ec:axes}), with anomalous $Wtb$ couplings, either real or purely imaginary. Here, for $P_x$ and $P_y$ the initial quark direction $\vec q$ is assumed known. }
\label{fig:Ptrue}
\end{center}
\end{figure*}

The variation of $P_i$ when a single anomalous coupling is non-zero is depicted in Fig.~\ref{fig:Ptrue}. The SM coupling $\vl$ is fixed to one from now on, and the range $[-0.2,0.2]$ chosen for the anomalous couplings is of the order of the current direct limits. The dependence of the top quark polarization on anomalous couplings can be extracted from a fit. The correction factors for the total single $t$ and single $\bar t$ cross sections with anomalous couplings are~\cite{AguilarSaavedra:2008gt}
\begin{eqnarray}
f_t & = & 1+0.90\, |\vR|^2+ 1.47\, |\gl|^2 + 2.31\, |\gr|^2  \notag \\
& &  -0.11\, \R\, \vl^* \vR - 0.53\, \R\, \vl^* \gr \,, \notag \\ 
f_{\bar t} & = & 1+ 1.09\, |\vR|^2+ 2.36\, |\gl|^2 + 1.58\, |\gr|^2  \notag \\
& & -0.12\, \R\, \vl^* \vR - 0.56\, \R\, \vR^* \gl \,,
\end{eqnarray}
where we have dropped
the combinations of couplings with numerical pre-factors smaller than 0.1. The polarizations are then
\begin{eqnarray}
P_x & = & \left( -0.13 |\vR|^2 + 0.25 \, |\gl|^2 -0.90\, |\gr|^2 \right. \notag \\
& & \left.  + 2.14\, \R\, \vl^* \gr -1.53 \,\R\, \vR^* \gl \right) / f_t \,, \notag \\
P_y & = & \left( -2.12\, \I\, \vl^* \gr -1.54 \,\I\, \vR^* \gl \right) / f_t \notag \\
P_z & = & \left( 0.90 - 0.76\, |\vR|^2 + 1.15\, |\gl|^2 - 1.50\, |\gr|^2 \right. \notag \\
& &  \left. -0.60\, \R\, \vl^* \gr + 0.36\, \R\, \vR^* \gl  \right) / f_t
\end{eqnarray} 
for top quarks and
\begin{eqnarray}
P_x & = & \left( -0.14 - 0.96 \, |\gl|^2 + 0.34\, |\gr|^2 \right. \notag \\
& & \left. - 1.71 \,\R\, \vl^* \gr +2.31 \,\R\, \vR^* \gl  \right) / f_{\bar t} \,, \notag \\
P_y & = & \left( 1.72\, \I\, \vl^* \gr +2.30 \,\I\, \vR^* \gl \right) / f_{\bar t} \notag \\
P_z & = & \left( -0.86 + 0.99\, |\vR|^2 - 1.56\, |\gl|^2 + 1.20\, |\gr|^2 \right. \notag \\
& &  \left. +0.42\, \R\, \vl^* \gr - 0.67\, \R\, \vR^* \gl  \right) / f_{\bar t}
\end{eqnarray} 
for antiquarks. The reason for the appearance of imaginary parts of coupling products in $P_y$, and not in $P_x$ nor $P_z$, is clear in the $2 \to 2$ approximation to the $t$-channel process. Since the differential cross section is proportional to the squared matrix element, and there are no absorptive parts, they can only arise from traces of Dirac matrices $\text{tr}\, \gamma^5 \gamma^\mu \gamma^\nu \gamma^\rho \gamma^\sigma = -4 i \epsilon^{\mu \nu \rho \sigma}$, with $\epsilon^{\mu \nu \rho \sigma}$ the totally antisymmetric tensor with $\epsilon^{0123}=1$, contracted with four different 4-vectors. In the $2 \to 2$ process there are only three independent 4-momenta, for example those of the initial and final state light quarks $p_q$, $p_j$ and the top quark $p_t$. Then, the fourth vector in a non-zero contraction must be the top spin vector $s_t$. In the top quark rest frame, the Lorentz-invariant
 contraction $\epsilon_{\mu \nu \rho \sigma} p_q^\mu p_j^\nu p_t^\rho s_t^\sigma$ is proportional to the triple product $(\vec p_q \times \vec p_j) \cdot \vec s_t$, which vanishes for $\vec s_t$ in the $\hat x$, $\hat z$ directions but not in the $\hat y$ direction. The extra $b$ quark in the $2 \to 3$ process affects little this argument, since it is mostly collinear to the beam direction.

\begin{figure}[htb]
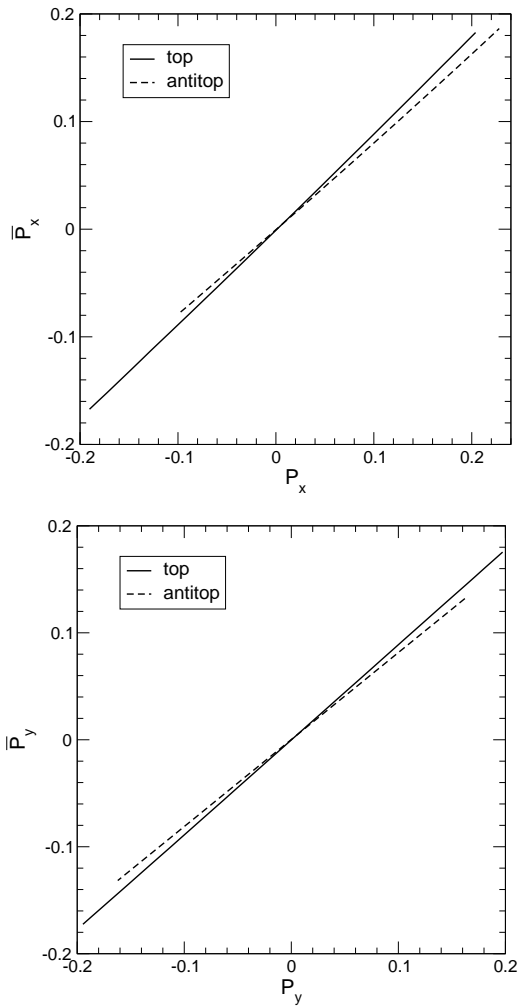

\begin{center}
\begin{tabular}{c}
\epsfig{file=fig2a.eps,height=6.5cm,clip=} \\[2mm]
\epsfig{file=fig2b.eps,height=6.5cm,clip=}  
\end{tabular}
\caption{Relation between the polarizations $P_{x,y}$, in which the initial quark direction is assumed known, and the polarizations $\bar P_{x,y}$, in which it is guessed (see the text).}
\label{fig:PvsP}
\end{center}
\end{figure}

As we have mentioned before, the direction of the initial quark in $qg \to q' t \bar b$ that allows to construct our $(x,y,z)$ reference system cannot be unambiguously determined. One has instead to choose between the two incoming proton beam directions by using some criterion -- the goodness of which will be the fraction of times that the true direction of the quark is correctly identified. To select among the directions of the two beams one can use the momentum of the spectator quark in the laboratory frame, which most times follows that of the initial quark. This choice gives the correct answer 97\% of the times for $ug \to d t \bar b$ and 98\% of the time for $dg \to u \bar t b$, which are the main channels for single top and antitop production, respectively. For the rest of channels the rate of correct estimations, which depends on the kinematics of the particular process, is smaller. This results in `observed' polarizations $\bar P_{x,y}$ that are slightly smaller than the `true' ones that one would measure if the quark direction were certainly known. The polarizations $\bar P_{x,y}$ can also be computed with a Monte Carlo. The relation between $P_{x,y}$ and $\bar P_{x,y}$ is linear, and independent of the anomalous couplings in the $[-0.2,0.2]$ intervals considered, as it can be observed in Fig.~\ref{fig:PvsP}. The ranges of the polarizations displayed are those that correspond to varying $\R\,\gr$ and $\I\,\gr$ between $-0.2$ and $0.2$, as in Fig.~\ref{fig:Ptrue}.
The numerical relation between $P_{x,y}$ and $\bar P_{x,y}$ is
\begin{align}
& \bar P_{x,y}= 0.89 P_{x,y}  && (t) \,, \notag \\
& \bar P_{x,y} = 0.81 P_{x,y}  && (\bar t) \,.
\end{align}
The numerical factor is the same for $P_x$ and $P_y$ because choosing either of the proton directions as the quark direction gives most of the times a minus sign in the normal and transverse directions. (Notice however that the two proton momenta are not anti-parallel in the top quark rest frame.)
For completeness, we also give here the results for the LHC with 14 TeV. 
The correction factors for the total cross sections with anomalous couplings are
\begin{eqnarray}
f_t & = & 1+0.92\, |\vR|^2+ 1.82\, |\gl|^2 + 2.60\, |\gr|^2  \notag \\
& &  -0.11\, \R\, \vl^* \vR - 0.47\, \R\, \vl^* \gr \,, \notag \\
f_{\bar t} & = & 1+ 1.07\, |\vR|^2+ 2.61\, |\gl|^2 + 1.92\, |\gr|^2   \notag \\
& & -0.12\, \R\, \vl^* \vR -0.12\, \R\, \vl^* \gr- 0.49\, \R\, \vR^* \gl \,, \notag \\
\end{eqnarray}
and the polarizations are
\begin{eqnarray}
P_x & = & \left(-0.10 |\vR|^2 -0.84\, |\gr|^2 + 0.31 |\gl|^2 
\right. \notag \\
& & \left. + 2.19\, \R\, \vl^* \gr   -1.68\,  \R\, \vR^* \gl \right) / f_t \,, \notag \\
P_y & = & \left(-2.17\, \I\, \vl^* \gr - 1.69 \,\I\, \vR^* \gl \right) / f_t   \,, \notag \\
P_z & = & \left( 0.88 - 0.77\, |\vR|^2 + 1.38\, |\gl|^2 - 1.71\, |\gr|^2 \right. \notag \\
& &  \left. -0.53\, \R\, \vl^* \gr + 0.35\, \R\, \vR^* \gl  \right) / f_t \,, \notag \\
\bar P_{x,y} & = & 0.88 P_{x,y} 
\end{eqnarray} 
for quarks and
\begin{eqnarray}
P_x & = & \left( -0.11 -0.88 |\gl|^2 +0.37\, |\gr|^2
\right. \notag \\
& & \left. - 1.84 \,\R\, \vl^* \gr + 2.31 \,\R\, \vR^* \gl \right) / f_{\bar t} \,, \notag \\
P_y & = & \left( 1.85\, \I\, \vl^* \gr + 2.31 \, \I\, \vR^* \gl \right) / f_{\bar t}  \,, \notag \\
P_z & = & \left( -0.85 + 0.95\, |\vR|^2 - 1.76\, |\gl|^2 + 1.43\, |\gr|^2 \right. \notag \\
& &  \left. +0.41\, \R\, \vl^* \gr - 0.601, \R\, \vR^* \gl  \right) / f_{\bar t} \,, \notag \\
\bar P_{x,y} & = & 0.79 P_{x,y}
\end{eqnarray} 
for antiquarks. 

\section{Polarization asymmetries}

We now come to discuss how to experimentally probe the top polarization.
It is well known that the distributions of the top quark decay products are sensitive to its polarization~\cite{Kuhn:1981md}. Let us consider for example the charged lepton, with three-momentum $\vec p_{\ell}$, whose direction in the $(x,y,z)$ reference system can be parameterised by the angles $(\theta_\ell,\phi_\ell)$. Integrating over $\phi_\ell$, the polar angle distribution is
\begin{equation}
\frac{1}{\Gamma} \frac{d\Gamma}{d\cos\theta_\ell} = \frac{1}{2} ( 1 + P_z \alpha_\ell \cos \theta_\ell ) \,,
\label{ec:dist1}
\end{equation}
where $\alpha_\ell$ is a constant named `spin analyzing power' of the lepton, which equals unity in the SM but in general depends on possible anomalous $Wtb$ couplings~\cite{AguilarSaavedra:2006fy}. 
(NLO corrections to this quantity are small~\cite{Bernreuther:2014dla}.) Then, a forward-backward (FB) asymmetry based on that angle
\begin{equation}
A_\text{FB}^z = \frac{\sigma(\cos \theta_\ell > 0) - \sigma(\cos \theta_\ell < 0)}{\sigma(\cos \theta_\ell > 0) + \sigma(\cos \theta_\ell < 0)} = \frac{1}{2} \alpha_\ell P_z
\end{equation}
\begin{figure}[htb]
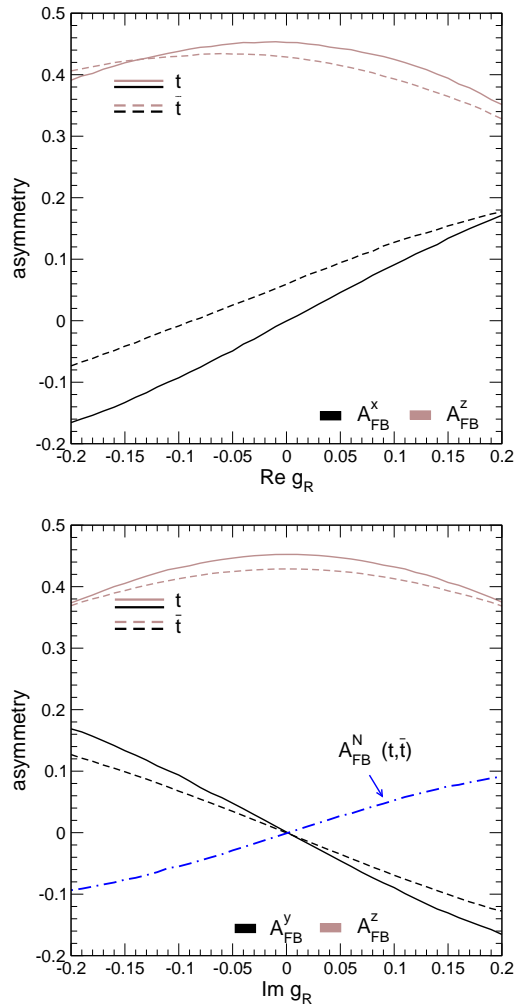

\begin{center}
\begin{tabular}{c}
\epsfig{file=fig3a.eps,height=6.5cm,clip=} \\[2mm]
\epsfig{file=fig3b.eps,height=6.5cm,clip=}  
\end{tabular}
\caption{Asymmetries sensitive to $\R\,\gr$ (up) and $\I\,\gr$ (down). In the bottom panel we also show for comparison the normal $W$ polarization asymmetry $A_\text{FB}^N$.}
\label{fig:AFB}
\end{center}
\end{figure}
is sensitive to $P_z$, and its measurement has actually been used by the CMS Collaboration to measure the top polarization -- under the assumption $\alpha_\ell=1$. The transverse and normal polarizations can be extracted from the azimuthal angle dependence of the double angular distribution $d\Gamma/d\cos \theta_\ell \,d\phi_\ell$.\footnote{In~\cite{Rindani:2011pk} the azimuthal distributions are investigated for $tW$ production, although the corresponding polarizations are not extracted.} But it is simpler to consider the angles between $\vec p_\ell$ and the $\hat x$, $\hat y$ directions, which lead to two more FB asymmetries,
\begin{equation}
A_\text{FB}^x = \frac{1}{2} \alpha_\ell \bar P_x \,,\quad A_\text{FB}^y = \frac{1}{2} \alpha_\ell \bar P_y \,,
\end{equation}
in the transverse and normal directions.
Neither of these three asymmetries provides a model-independent measurement of the top quark polarization, as new physics in the decay vertex can enter the distributions through the $\alpha_\ell$ factor. (In contrast, a model-independent measurement would be possible by the analysis of double angular distributions also involving $W$ rest-frame momenta~\cite{AguilarSaavedra:2012xe}.) However, if we are interested only in new physics yielding anomalous $Wtb$ couplings, and not other dimension-six operators such as four-fermion ones~\cite{Cao:2007ea,Zhang:2010dr,AguilarSaavedra:2010zi}, limits on the anomalous couplings can be extracted from the measurement of these asymmetries, especially of $A_\text{FB}^x$ and $A_\text{FB}^y$.
Their respective dependence on $\R\,\gr$ and $\I\,\gr$, given through the polarizations $P_i$ and the electron spin analyzing power $\alpha_\ell$, is depicted in Fig.~\ref{fig:AFB}.
In the lower panel we also plot the dependence of the normal $W$ polarization asymmetry $A_\text{FB}^N$~\cite{AguilarSaavedra:2010nx} on $\I\,\gr$. This asymmetry has recently been measured by the ATLAS Collaboration, $A_\text{FB}^N =  0.031 \pm 0.065\;\text{(stat)}^{+0.029}_{-0.031}\;\text{(syst)}$~\cite{ATLAS:2013ula}  using 4.7 fb$^{-1}$ at 7 TeV and provides the first direct limit on $\I\,\gr$.

By comparing the dependence on $\gr$ it is clear that the new asymmetries $A_\text{FB}^x$ ($A_\text{FB}^y$) have a stronger dependence on $\R\,\gr$ ($\I\,\gr$) than the asymmetry $A_\text{FB}^z$ measured by the CMS Collaboration. Moreover, systematic uncertainties on $A_\text{FB}^{x,y}$ should be smaller, since their SM value is zero, except in the case of $A_\text{FB}^x$ for $\bar t$ production. (One can compare with the CMS measurement $A_\text{FB}^z = 0.41 \pm 0.06\;\text{(stat)} \pm 0.16 \;\text{(sys)}$ that has a relatively large $\sim 40\%$ systematic uncertainty.) We can reasonably expect that the experimental uncertainties on $A_\text{FB}^{x,y}$ and $A_\text{FB}^N$ will all be of similar magnitude, since all these asymmetries vanish in the SM, except $A_\text{FB}^x$ for $\bar t$, and their measurement requires full kinematical reconstruction of the single top events, the main difference being that in $A_\text{FB}^N$ the charged lepton distribution in the $W$ boson rest frame is considered. If this is indeed the case, the sensitivity of the normal asymmetry $A_\text{FB}^y$ to $\I\,\gr$ is a factor of 1.5 higher: from Fig.~\ref{fig:AFB}, for small $\gr$ one approximately has 
\begin{align}
& A_\text{FB}^y = -0.95 \, \I\,\vl^* \gr && (t) \,,\notag \\
& A_\text{FB}^y = -0.67 \, \I\, \vl^* \gr && (\bar t) \,,\notag \\
& A_\text{FB}^N = 0.54 \,  \,\I\, \vl^* \gr && (t,\bar t) \,.
\end{align}
We can conservatively assume that the systematic uncertainties at 8 TeV are the same as in the current measurement~\cite{ATLAS:2013ula}, because the evaluation of systematic uncertainties usually benefits from larger data and Monte Carlo samples. In this case, the attainable limit will be $\I\,\gr \in[-0.06,0.06]$, far better than the current one $\I\, \gr \in[-0.3,0.4]$. From the measurement of the transverse asymmetry $A_\text{FB}^x$, a limit $\R\,\gr \in[-0.06,0.06]$ is also expected. This sensitivity is competitive with the current limit $\R\, \gr \in[-0.08,0.04]$ from asymmetries in $t \bar t$ production, which is already systematics dominated~\cite{Aad:2012ky}. We also note that the comparison of our limits with the ones in~\cite{Rindani:2011pk} for the $tW$ process
 is not easy, since those are obtained without taking systematic uncertainties into account, and the measurements we consider are dominated by systematics given the large data samples available at the LHC.

\section{Summary}

In this paper we provide a complete description of the top (anti-)quark polarization in single top production, by introducing two directions, transverse and normal, which are orthogonal to the already used spectator quark direction. Given the polarizations in these three axes, $P_x$, $P_y$ and $P_z$, the top (anti)quark spin density matrix can be determined and thus the polarization in any other direction can be computed.

For single top quark production the polarizations $P_x$, $P_y$ newly defined vanish in the SM, while for antiquarks $P_x$ is of the order of $\mathcal{O}(0.1)$, with $P_y$ still vanishing. This fact makes these polarizations very sensitive to new physics contributions, such as top anomalous couplings or four-fermion operators. We have focused on the former, giving
the dependence of the polarizations on possible anomalous $Wtb$ couplings. It has been shown that two asymmetries, involving the transverse ($P_x$) and normal ($P_y$) polarizations, are very sensitive to an anomalous coupling $\gr$ involving a dipole term of the form $\bar b_L \sigma^{\mu \nu} t_R$. Their measurement at the LHC will likely improve the limits on $\gr$. The expected limits on the real part of this coupling with 8 TeV data are slightly better than the current ones at 7 TeV, which are already systematics-dominated. The expected limits on its imaginary part should improve the current ones by a factor of 5. And, in any case, limits from these new asymmetries can be combined with previously known observables in order to improve the sensitivity.

\acknowledgements
We thank R.V. Herrero-Hahn for collaboration at the initial stages of this work.
This work has been supported by MICINN project FPA2010-17915; by
Junta de Andaluc\'{\i}a projects FQM 101, FQM 03048 and FQM 6552; and by FCT project EXPL/FIS-NUC/0460/2013. The work of S. Amor dos Santos was also supported by FCT grant SFRH/BD/73438/2010.


\begin{thebibliography}{99}

\bibitem{Mahlon:1996pn} 
  G.~Mahlon and S.~J.~Parke,
  Phys.\ Rev.\ D {\bf 55}, 7249 (1997)
  [hep-ph/9611367].

\bibitem{Mahlon:1999gz} 
  G.~Mahlon and S.~J.~Parke,
  Phys.\ Lett.\ B {\bf 476}, 323 (2000)
  [hep-ph/9912458].

\bibitem{CMS:2013rfa} 
  S. Chatrchyan {\it et al.} [CMS Collaboration],
  CMS-PAS-TOP-13-001.

\bibitem{Espriu:2001vj} 
  D.~Espriu and J.~Manzano,
  Phys.\ Rev.\ D {\bf 65}, 073005 (2002)
  [hep-ph/0107112].

\bibitem{AguilarSaavedra:2008gt} 
  J.~A.~Aguilar-Saavedra,
  Nucl.\ Phys.\ B {\bf 804}, 160 (2008)
  [arXiv:0803.3810 [hep-ph]].

\bibitem{cteq6l1}
  P.~M.~Nadolsky, H.~-L.~Lai, Q.~-H.~Cao, J.~Huston, J.~Pumplin, D.~Stump, W.~-K.~Tung and C.~-P.~Yuan,
  Phys.\ Rev.\ D {\bf 78} (2008) 013004
  [arXiv:0802.0007 [hep-ph]].

\bibitem{Schwienhorst:2010je} 
  R.~Schwienhorst, C.~-P.~Yuan, C.~Mueller and Q.~-H.~Cao,
  Phys.\ Rev.\ D {\bf 83}, 034019 (2011)
  [arXiv:1012.5132 [hep-ph]].

\bibitem{Campbell:2009ss} 
  J.~M.~Campbell, R.~Frederix, F.~Maltoni and F.~Tramontano,
  Phys.\ Rev.\ Lett.\  {\bf 102}, 182003 (2009)
  [arXiv:0903.0005 [hep-ph]];
  JHEP {\bf 0910}, 042 (2009).
  [arXiv:0907.3933 [hep-ph]].


\bibitem{AguilarSaavedra:2008zc} 
  J.~A.~Aguilar-Saavedra,
  Nucl.\ Phys.\ B {\bf 812}, 181 (2009)
  [arXiv:0811.3842 [hep-ph]].



\bibitem{AguilarSaavedra:2006fy} 
  J.~A.~Aguilar-Saavedra, J.~Carvalho, N.~F.~Castro, A.~Onofre and F.~Veloso,
  Eur.\ Phys.\ J.\ C {\bf 50}, 519 (2007)
  [hep-ph/0605190].

\bibitem{Aad:2012ky} 
  G.~Aad {\it et al.}  [ATLAS Collaboration],
  JHEP {\bf 1206}, 088 (2012)
  [arXiv:1205.2484 [hep-ex]].

\bibitem{AguilarSaavedra:2011ct} 
  J.~A.~Aguilar-Saavedra, N.~F.~Castro and A.~Onofre,
  Phys.\ Rev.\ D {\bf 83}, 117301 (2011)
  [arXiv:1105.0117 [hep-ph]].



\bibitem{Kane:1991bg} 
  G.~L.~Kane, G.~A.~Ladinsky and C.~P.~Yuan,
  Phys.\ Rev.\ D {\bf 45}, 124 (1992).

\bibitem{CMS:2013pfa} 
  S. Chatrchyan {\it et al.}  [CMS Collaboration],
  CMS-PAS-TOP-13-008.


\bibitem{AguilarSaavedra:2010nx} 
  J.~A.~Aguilar-Saavedra and J.~Bernabeu,
  Nucl.\ Phys.\ B {\bf 840}, 349 (2010)
  [arXiv:1005.5382 [hep-ph]].

\bibitem{ATLAS:2013ula} 
  G. Aad {\it et al.} [ATLAS Collaboration],
  ATLAS-CONF-2013-032.

\bibitem{Grzadkowski:2008mf} 
  B.~Grzadkowski and M.~Misiak,
  Phys.\ Rev.\ D {\bf 78}, 077501 (2008)
  [Erratum-ibid.\ D {\bf 84}, 059903 (2011)]
  [arXiv:0802.1413 [hep-ph]].

\bibitem{Drobnak:2011wj} 
  J.~Drobnak, S.~Fajfer and J.~F.~Kamenik,
  Phys.\ Lett.\ B {\bf 701}, 234 (2011)
  [arXiv:1102.4347 [hep-ph]].

\bibitem{Kuhn:1981md} 
  J.~H.~Kuhn and K.~H.~Streng,
  Nucl.\ Phys.\ B {\bf 198}, 71 (1982).

\bibitem{Bernreuther:2014dla} 
  W.~Bernreuther, P.~Gonz\'alez and C.~Mellein,
  arXiv:1401.5930 [hep-ph].

\bibitem{Rindani:2011pk} 
  S.~D.~Rindani and P.~Sharma,
  JHEP {\bf 1111}, 082 (2011)
  [arXiv:1107.2597 [hep-ph]].

\bibitem{AguilarSaavedra:2012xe} 
  J.~A.~Aguilar-Saavedra and R.~V.~Herrero-Hahn,
  Phys.\ Lett.\ B {\bf 718}, 983 (2013)
  [arXiv:1208.6006 [hep-ph]].

\bibitem{Cao:2007ea} 
  Q.~-H.~Cao, J.~Wudka and C.~-P.~Yuan,
  Phys.\ Lett.\ B {\bf 658}, 50 (2007)
  [arXiv:0704.2809 [hep-ph]].

\bibitem{Zhang:2010dr} 
  C.~Zhang and S.~Willenbrock,
  Phys.\ Rev.\ D {\bf 83}, 034006 (2011)
  [arXiv:1008.3869 [hep-ph]].

\bibitem{AguilarSaavedra:2010zi} 
  J.~A.~Aguilar-Saavedra,
  Nucl.\ Phys.\ B {\bf 843}, 638 (2011)
  [Erratum-ibid.\ B {\bf 851}, 443 (2011)]
  [arXiv:1008.3562 [hep-ph]].

\end{thebibliography}
\end{document}